\documentclass[aps,prb,showpacs,superscriptaddress, twocolumn, footinbib]{revtex4-1}

\usepackage{amsmath}
\usepackage{graphicx}
\usepackage{dcolumn}
\newcolumntype{d}[1]{D{.}{.}{#1}}

\usepackage{bm}
\usepackage{epstopdf}
\usepackage{graphicx}
\usepackage{stmaryrd}
\usepackage{tikz}
\usepackage{lipsum}
\usepackage{pgf}
\usepackage{float}
\usepackage{hyperref}
\usepackage{tabularx}
\usepackage{appendix}
\usepackage[section]{placeins}
\usepackage{subfigure}

\definecolor{lightblue}{RGB}{0,170,255}

\newcommand{\appropto}{\mathrel{\vcenter{
			\offinterlineskip\halign{\hfil$##$\cr
				\propto\cr\noalign{\kern2pt}\sim\cr\noalign{\kern-2pt}}}}}

\usepackage[normalem]{ulem}
\newcommand{\dps}{\displaystyle}

\newcommand{\om}{\iffalse}
\newcommand{\pd}[2]{\frac{\partial #1}{\partial #2}}

\newcommand{\ba}{\arraycolsep 0.3ex \begin{array}{rl}}
	\newcommand{\ea}{\end{array}}
\newcommand{\bc}{\begin{cases}}
	\newcommand{\ec}{\end{cases}}
\usepackage{color}
\newcolumntype{C}[1]{>{\centering\arraybackslash}p{#1}}

\begin{document}
	\title{Intrinsic Contribution to Non-linear Thermoelectric Effect}

\author{Pankaj Bhalla}
\affiliation{Beijing Computational Science Research Center, Beijing, 100193, China}

\date{\today}
\begin{abstract}
Irradiation of the strong light on the material leads to numerous non-linear effects that are essential to understand the physics of excited states of the system and for optoelectronics. Here, we study the non-linear thermoelectric effect due to the electric and thermal fields applied on a non-centrosymmetric system. The phenomenon arises on the Fermi surface with the transitions of electrons from valence to conduction bands. We derive the formlism to investigate these effects and find that the non-linearity in these effects namely non-linear Seebeck and non-linear Peltier effects depends on the ratio of the non-linear to the linear conductivities. The theory is tested for a hexagonally warped and gapped topological insulator. Results show enhancement in the longitudinal and Hall effects on increasing the warping strength while show opposite behavior with the surface gap.
\end{abstract}	
\maketitle

\section{Introduction}
A road in pursuit of highly efficient thermoelectric materials is challenging and fundamentally important as directly relates with the global energy crisis issue. This makes indespensable for the researchers to puzzle out the thermoelectric transport properties \cite{alam_NE2013}. These properties decide the thermal to electrical energy conversion efficiency, governed by the dimensionless quantity named as figure of merit $zT = \sigma S^2T/\kappa$, \cite{rowe_book, behnia_book} where $T$ is an absolute temperature, $\sigma$ is an electrical conductivity, $S$ is a Seebeck coefficient and $\kappa = \kappa_e + \kappa_{\text{lattice}}$ is the sum of electrical ($\kappa_e$) and lattice ($\kappa_{\text{lattice}}$) thermal conductivities. Inovative approaches and concepts have been employed for decades to enhance the performance of such materials and hierarchical architecturing.\cite{koumoto_book, dresselhaus_AdMat2007, poudel_Sci2008, sootsman_ACIE2009, tretiakov_APL2011, pei_nature2011, biswas_nature2012, osterhage_APL2014, Li_CPL2015, Xu_CPB2016, xu_npjQM2017, baldomir_SR2019}\\
Approaches deal with the responses to the first power of external field provides great success\cite{gayner_PMS2016}, if the amplitude of the incident light is small. However if the amplitude becomes sufficiently large to distort electron's orbits, it becomes crucial to look after the responses to higher order fields to generate new optoelectronic devices and new quantum technologies. In the following years, a lot of work has been done to study the non-linear electrical effects such as shift current, second-harmonic generation, injection current and anomalous current, resonant photovoltaic current \cite{kraut_PRB1981, Khurgin_JOS1994, Khurgin_APL1995, Sipe_PRB2010, rappe_12, Inglot_PRB2015, shilie_16, wenhui_17, tokura_17, Kim_Shift_PRB2017, alexey_16, rappe_16, nagaosa_16,Khurgin_JOSAB2016, nagaosa_16a, morimoto_PRB2016, morimoto_SA2016, Hamamoto_NLSC_PRB2017, Moore_NATComm2017, benjamin_PRB2017, NLH_Tune_2DM2018, krasnok_MT2018, tan_PRB2019, zhang_2019, benjamin_PRB2019, yinan_NM2019, bhalla_PRL2020, sekine_PRB2020, ahn_arxiv2020, wang_arxiv2020}. But only few work has been devoted to the thermoelectric effects \cite{rosa_PRB2013, hwang_PRB2014, Gaoyang_PRB2018}. To fill the gap and for better understanding about the non-linear behavior of thermoelectric properties, we provide a theory for non-linear thermoelectric conductivities and the related phenomenon such as Seebeck effect, Peltier effect in the weakly non-linear regime. This has been examined for hexagonally warped and gapped topological insulators \cite{Fu_PRL2009}.\\
\begin{figure}[htbp]
\centering
\hspace{-0.8cm}
\includegraphics[width=9cm, height=4cm]{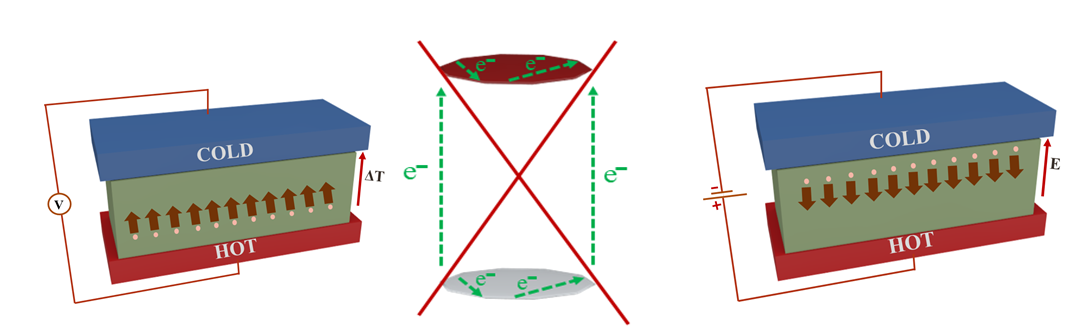}
\caption{Schematic view for the generation of thermoelectric effects due to interband and intraband electronic transitions shown in the middle picture with green dashed lines. Left diagram refers to the Seebeck effect and Right to the Peltier effect and middle corresponds to the energy spectrum around Dirac point having distorted Fermi surface.}
\label{fig:RatioAT}
\end{figure}
Basically, the transport properties are governed by two different compartments with one belongs to the applied field or known as field driving term and other to scattering processes. Former part plays a significant role in the dissipationless transport phenomenon and relates with the momentum dependent Bloch wave functions. More specifically, the blend of interband contribution, characterized by Berry connection that relates two neighboring momentum dependent wave functions and intraband contribution due to the excitation of carriers within the same band in response to the external field leads to give intrinsic effects \cite{culcer_17}. In systems having time-reversal symmetry or inversion symmetery breaking, well known examples such as anomalous Hall effect \cite{nagaosa_RMP2010}, chiral magnetic effect \cite{Burkov_PRB2012, monteiro_PRB2015, Shuang_NatMat2016, kaushik_PRB2019} and negative magnetoresistance in 3D Dirac and Weyl semimetals \cite{li_NComm2016, schumann_PRB2017, armitage_RMP2018, culcer_2DM2020} has been observed due to such consequences. The latter effects due to the interactions between carriers like electron-impurity, charged impurities, phonons, etc. are known as extrinsic effects. Here we are mainly focus on the contribution made by intrinsic effects to thermoelectric properties. We find that for non-centrosymmetric systems, the deviation of thermoelectric effects from linear regime depends on the ratio of the non-linear to the linear responses.\\
To exemplify our theory, we consider time-reversal invariant topological insulator $Bi_2Te_3$ having nontrivial topological order\cite{Fu_PRL2009, moore_nature2010, hasan_ARCMP2011, chen_PRB2011, pollmann_PRB2012}. We observe that the responses relies on the Bloch band dispersion near the Fermi surface of this material, hence get attention by thermoelectric community \cite{xu_npjQM2017, matsushita_PRM2017, ivanov_PSS2018, wang_NE2019}. Detailed investigation yields vanishing first-order Hall and second-order longitudinal response due to intraband transitions. However, it produces non-zero first-order longitudinal conductivity and second-order Hall conductivity. Due to the transitions between valence and conduction bands, the first-order Hall and second-order longitudinal response gives finite value. Here the responses depend on the degree of distortion and the strength of gap value.\\
This paper is organized as follows. In Sec.~I, we provide the detailed theoretical framework to compute the thermoelectric effect. First, we give the quantum kinetic theory and the basic structure for the generalized currents. Later we express the non-linear Seebeck and Peltier effects in terms of the ratio of conductivities of different order to external fields. In next section, we apply our theory to the modelled Hamiltonian for Bi$_2$Te$_3$ and demonstrate the results for electrical and thermal currents in response to the applied fields. Finally, we summarize our study with future implications.
  
\section{Theoretical Framework}
We embark our calculation with the quantum Liouville equation for the density matrix \cite{culcer_17}
\begin{equation}
\ba
&\dps\frac{d\rho}{dt} + \frac{i}{\hbar} [H,\rho] = 0,
\label{eqn:QLE}
\ea
\end{equation}
with $H$ the total Hamiltonian, a sum of unperturbed and perturbed parts. The unperturbed part is the band Hamiltonian $H_0$ and $H_{\text{pert}} = H_{F} + U$ having $H_F$ as an interaction with external fields such as electric, thermal, and magnetic and $U$ as the interaction potential. Substituting this into Eq.~(\ref{eqn:QLE}) and expanding the density matrix perturbatively into the powers of field such as $\rho = \rho^{(0)} + \rho^{(1)} + \cdots $, where superscripts indicate the order in the field, Eq.~(\ref{eqn:QLE}) becomes
\begin{equation}
\ba
&\dps \frac{d\rho^{(n)}}{dt} + \frac{i}{\hbar} [H_0, \rho^{(n)}] + J(\rho^{(n)}) = \mathcal{D}_{F}(\rho^{(n-1)}).
\label{eqn:rhon}
\ea
\end{equation}
Here $\mathcal{D}_{F} = -i/\hbar [H_F,\rho^{(n-1)}]$ is the field driving term and $J(\rho^{(n)})$ is the scattering term which in the Born approximation takes the form
\begin{equation}
\ba
J(\rho^{(n)}) &\dps = \frac{1}{\hbar^2} \, \int^\infty_0 dt' \, \langle[U, [e^{-\frac{iH_0t'}{\hbar}}U e^{\frac{iH_0t'}{\hbar}}, \rho^{n}]]\rangle.
\ea
\end{equation}
To solve the kinetic equation, we segregate the density matrix into diagonal, $f_{d,{\bm k}}$ and off-diagonal, $f_{\text{od},{\bm k}}$ components using $\rho^{(n)} = f_{d,{\bm k}}^{(n)} + f_{\text{od},{\bm k}}^{(n)}$ and express the Liouville equation (Eq.~\ref{eqn:rhon}) into two coupled equations. Here the diagonal part accounts to the intraband and off-diagonal part to the interband coherence effects. This coupling part stems from the scattering term $J$, a sum of four parts such as $J = J_{d}(f_{d,{\bm k}}^{(n)}) + J_{d}(f_{\text{od},{\bm k}}^{(n)}) + J_{\text{od}}(f_{d,{\bm k}}^{(n)}) + J_{\text{od}}(f_{\text{od},{\bm k}}^{(n)})$.
\begin{equation}
\ba
&\dps \frac{d f_{d,{\bm k}}^{(n)}}{dt} + \frac{i}{\hbar} [H_0, f_{d,{\bm k}}^{(n)}] + J_{d}[f_{d,{\bm k}}^{(n)}] = \mathcal{D}_{F}(f_{d,{\bm k}}^{(n-1)}) - J_{d}[f_{\text{od},{\bm k}}^{(n)}],
\label{eqn:rhodiagonaln}
\ea
\end{equation}
and
\begin{equation}
\ba
&\dps \frac{d f_{\text{od},{\bm k}}^{(n)}}{dt} + \frac{i}{\hbar} [H_0, f_{\text{od},{\bm k}}^{(n)}] + J_{\text{od}}[f_{\text{od},{\bm k}}^{(n)}] = \mathcal{D}_{F}(f_{\text{od},{\bm k}}^{(n-1)}) - J_{\text{od}}[f_{\text{d},{\bm k}}^{(n)}].
\label{eqn:rhooffdiagonaln}
\ea
\end{equation}
The transport calculations has been performed in response to the temperature gradient within the relaxation time approximation\cite{bhalla_PRL2020}. Under this assumption, we consider $J_d(f_{\text{d}},{\bm k}^{(n)}) = f_{\text{d}},{\bm k}^{(n)}/\tau_d$, and $J_{\text{od}}(f_{\text{od}},{\bm k}^{(n)}) = f_{\text{od}},{\bm k}^{(n)}/\tau_{\text{od}}$ with $\tau$ a relaxation time. For convenience, we take $\tau_\text{d}$ and $\tau_{\text{od}}$ as same parameter and treat as a constant across the Fermi surface. With this, the solutions for diagonal and off-diagonal parts of density matrix become
\begin{equation}
\ba
f_{\text{d},{\bm k}}^{(n)} &\dps = \frac{\tau}{\hbar} \frac{{\bm \nabla} T}{T} \cdot \pd{f_{\bm k}^{(n-1)}}{{\bm k}} (\varepsilon_{\bm k} ^{m} - \mu) - \tau J_{d}[f_{\text{od},{\bm k}}^{(n)}],\\[2ex]
f_{\text{od},{\bm k}}^{(n),mm'} &\dps = -i\hbar \frac{[\mathcal{D}(\rho^{(n-1)}_{\bm k})]^{mm'} - [J(f_{d,{\bm k}}^{(n)})]^{mm'}}{\varepsilon_{\bm k}^{m} - \varepsilon_{k}^{m'}}.
\label{eqn:density}
\ea
\end{equation}  
Here $\varepsilon_{\bm k}$ a band dispersion, $\mu$ a chemical potential, and the thermal driving term $\mathcal{D}(\rho^{(n-1)}_{\bm k}) = \frac{1}{\hbar}\frac{{\bm \nabla} T}{T} \cdot \frac{D (\{H_0,\rho^{(n-1)}\}}{D{\bm k}}$ with $\{ ,\}$ an anticommutator and $\frac{D A}{D{\bm k}} = \pd{A}{{\bm k}} - i [\mathcal{R}_{{\bm k}},A]$ a covariant derivative with respect to the wave vector and $\mathcal{\bm R}_{\bm k}^{mm'} = \langle u_{\bm k}^{m}\vert i \partial_{\bm k} u_{\bm k}^{m'}\rangle$ a Berry connection with $\vert u_{\bm k}^{m} \rangle$ a momentum dependent quantum Bloch wave function having band index $m$, and $A$ is an arbitrary matrix. Due to the absence of a commutator, there is no covariant derivative for the diagonal part. The detailed derivation of Eq.~(\ref{eqn:density}) is mentioned in Appendix. 
\subsection{Generalized currents}
In the presence of field, the current can be expanded in the powers of the applied external field. Schematically, it can be expressed like
\begin{equation}
\ba
j_{a} &\dps = \sum_{b}A_{ab}F_{b} + \sum_{b,c}A_{abc} F_{b}F_{c} + \sum_{b,c,d}A_{abcd} F_{b} F_{c} F_{d} + \cdots ,
\label{eqn:gencurrent}
\ea
\end{equation}
where $F$ is an external field in an arbitrary direction indicated by subscript, $A_{ab}$ and $A_{abc}$ are second- and third-rank tensors respectively. The first term is the well known linear response, and the higher order terms are non-linear responses. The even-order terms only contribute to noncentrosymmetric systems having broken inversion symmetry such as ferroelectrics \cite{Fridkin_book}. In other systems, these terms vanish. In this work, we will consider systems having breaking inversion symmetry to calculate the conductivity tensor in response to the applied field. Here we assume small external field, thus consider response upto the second-order case.\\
From the basic definition, the electric and thermal currents are \cite{ziman_book}
\begin{equation}
\ba
{\bm J}^E &\dps= \frac{-e}{\hbar}\sum_{\bm k}{\bm v}_{\bm k} \rho_{{\bm k}}\\[2ex]
{\bm J}^Q &\dps= \frac{1}{2\hbar}\sum_{\bm k}(\varepsilon_{\bm k} - \mu) {\bm v}_{\bm k} \rho_{{\bm k}}.
\label{eqn:currentdef}
\ea
\end{equation}
Here ${\bm v}_{\bm k}$ is a $2 \times 2$ matrix and represented as
\begin{equation}
\ba
{\bm v}_{\bm k} &\dps= \left(
\begin{array}{cc}
\partial_{\bm k}\varepsilon_{\bm k}^{m} &   -i\mathcal{R}_{\bm k}^{mm'} (\varepsilon_{\bm k}^{m} - \varepsilon_{\bm k}^{m'})   \\
-i\mathcal{R}_{\bm k}^{m'm} (\varepsilon_{\bm k}^{m} - \varepsilon_{\bm k}^{m'}) & \partial_{\bm k}\varepsilon_{\bm k}^{m'}
\end{array}
  \right).
\ea
\end{equation}
Using the above matrix for velocity and the density matrix, the thermal and electrical currents are lead by three terms, mentioned below 
\begin{equation}
\ba
j_1 &\dps = \partial_{\bm k}\varepsilon_{\bm k}^{m} f_{d,{\bm k}}^{m},\\[2ex]
j_2 &\dps = -Re[\mathcal{R}_{\bm k}^{mm'}] (\varepsilon_{\bm k}^{m} - \varepsilon_{\bm k}^{m'}) Im[f_{\text{od},{\bm k}}^{mm'}] ,\\[2ex]
j_3 &\dps = Im[\mathcal{R}_{\bm k}^{mm'}] (\varepsilon_{\bm k}^{m} - \varepsilon_{\bm k}^{m'}) Re[f_{\text{od},{\bm k}}^{mm'}].
\ea
\end{equation}
Last two terms survive only for $m\neq m'$, thus describe the interband coherence effects to responses. In the next section, we calculate the nonlinear thermoelectric effect using these definitions.
\subsection{Nonlinear Thermoelectric effects}
The conversion of electric voltage to temperature differences and vice-versa lead to thermoelectric effects such as Seebeck effect, Peltier effect and Thomson effect \cite{behnia_book, goldsmid_book2017}. The phenomenon to generate the potential difference across the material due to the diffusion of charge carriers along the temperature gradient is known as Seebeck coefficient, denoted by $S$. 
\begin{equation}
\ba
S &\dps = \frac{J_Q^E}{J_E^E},
\ea
\end{equation}
where $J_Q^E$ is the thermal current induced by the electric field and $J_E^E$ is the electric current. \\
The reverse phenomenon of the former effect means the generation of flow of heat due to the applied electric potential is known as Peltier effect, denoted by $\Pi$.
\begin{equation}
\ba
\Pi &\dps = \frac{J_Q^Q}{J_E^E},
\ea
\end{equation}
having $J_Q^Q$ as the thermal current generated by thermal field. Moreover, the above expressions are interconnected by Kelvin relation $\Pi = T S$. However, these expressions are established within the linear response regime. We generalize these expressions to higher orders in field. Using the defintion of generalized current (Eq.~\ref{eqn:gencurrent}) the above relations can be written like
\begin{equation}
\ba
S &\dps = \frac{\alpha_1  + \alpha_2 {\bm \nabla} T + \cdots}{\sigma_1   + \sigma_2 {\bm E} + \cdots},
\ea
\end{equation}
with $\alpha_i$, and $\sigma_i$ are $i^{\text{th}}$-order thermoelectric and electrical conductivity tensors, ${\bm E}$ and ${\bm \nabla}T$ are electric field and temperature gradient respectively.\\
In the weakly nonlinear regime, we expand the denominator and keep the terms to the first power of field. Thus, we have
\begin{equation}
\ba
&\dps S  = \frac{\alpha_1}{\sigma_1} + \frac{\alpha_2}{\sigma_1}\nabla T - \frac{\alpha_1\sigma_2}{\sigma_1^2} E \cdots
\ea
\end{equation} 
On substituting $S_0 = \alpha_1 / \sigma_1$ corresponding to the linear case, the effect can be expressed like 
\begin{equation}
\ba
&\dps S  = S_0 \bigg( 1 + \frac{\alpha_2}{\alpha_1}\nabla T - \frac{\sigma_2}{\sigma_1} E \cdots \bigg).
\label{eqn:SE}
\ea
\end{equation} 
Similarly, the Pelteir effect in non-linear regime is
\begin{equation}
\ba
&\dps \Pi  = \Pi_0 \bigg( 1 + \frac{\kappa_2}{\kappa_1}\nabla T - \frac{\sigma_2}{\sigma_1} E \cdots \bigg),
\label{eqn:PE}
\ea
\end{equation}
having $\kappa_i$ an $i^{\text{th}}$-order thermal conductivity tensor and $\Pi_0 = \kappa_1/\sigma_1$.\\
From above expressions, it is evident that the ratio of the conductivities such as the non-linear to the linear decides the conversion efficiency of currents.\\
In the following subsections, we test our theoretical approach with the model Hamiltonian and calculate the electric and thermal currents due to electric and thermal gradient fields. 

\subsubsection{Model Hamiltonian}
We consider a band Hamiltonian describing the surface states of the topological insulators\cite{Fu_PRL2009, Murkani_PRB2011, xiao_PRB2013, Chang_PRB2015, Akzyanov_PRB2018}.
\begin{equation}
\ba
H_0 &\dps= v_F (k_x \sigma_y - k_y \sigma_x) + \Delta \sigma_z + \frac{\lambda}{2} (k_{+}^{3} + k_{-}^{3})\sigma_z,
\label{eqn:bandHam}
\ea
\end{equation}
where $v_F$ is the Fermi velocity, $\Delta$ is the surface gap, $\lambda$ is the hexagonal warping coefficient, $\sigma_i$ are Pauli matrices, $k_{\pm} = k_x \pm i k_y$ with $k_x$ ($k_y$) as x-component (y-component) momentum of the quasiparticle. The energy spectrum associated with the Hamiltonian is
\begin{equation}
\ba
\varepsilon_{{\bm k},\pm} &\dps = \pm \sqrt{v_F^2 k^2 + (\lambda k^3 \cos 3\theta + \Delta)^2},
\ea
\end{equation}
having $\theta$ a polar angle to define the momentum direction in the two dimensional surface state and $\pm$ represents band index.

\begin{figure}[htbp]
\centering
\includegraphics[width=14cm,height=8cm]{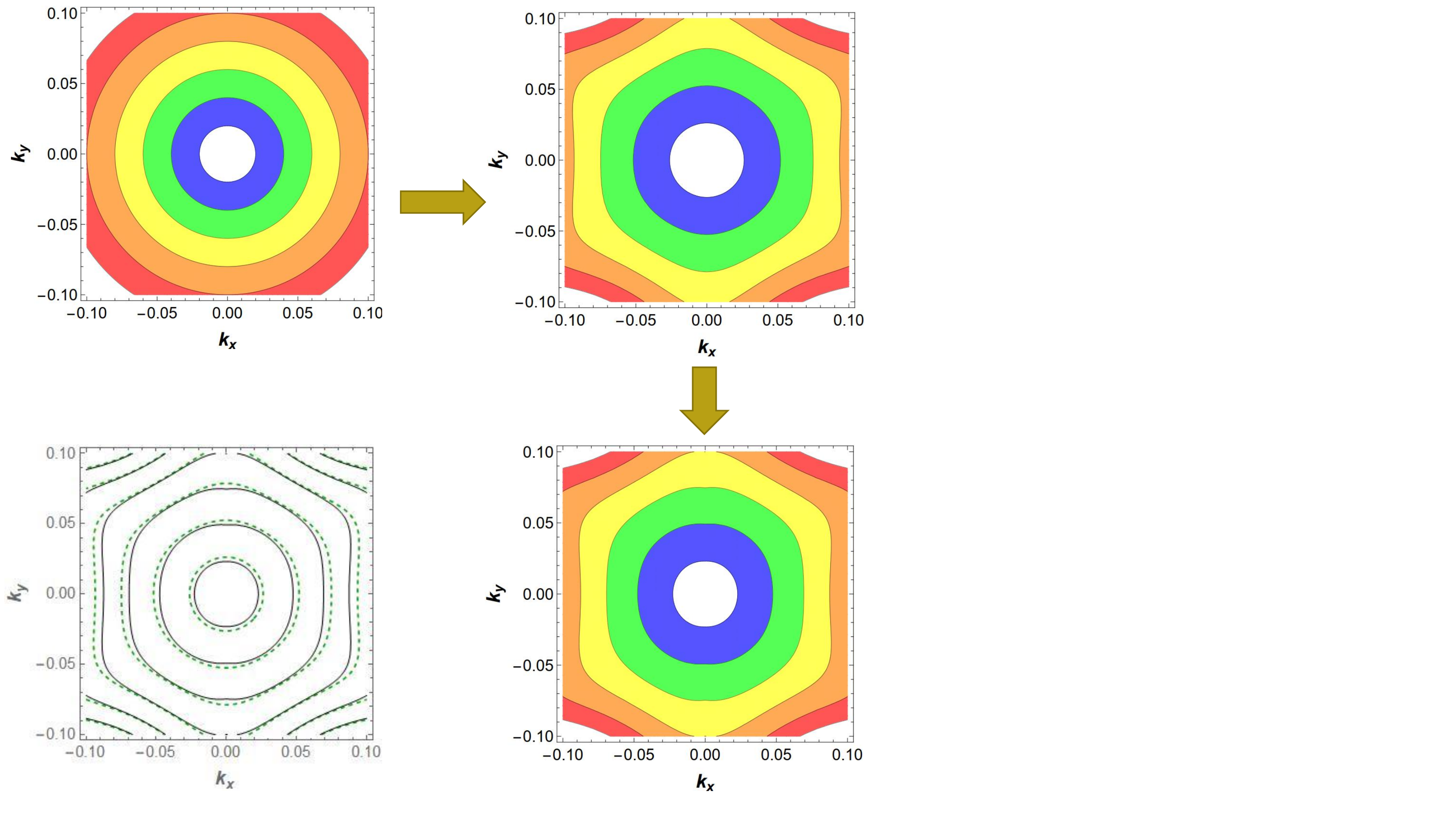}
\caption{Constant energy contours to describe the surface structure of a topological insulator. Top left: with Rashba spin-orbit effect, Top right: Rashba and Warping effect, Bottom right: Rashba, Warping and gap effect, Bottom left: the solid black curves are for the surface gap $\Delta = 0$ eV and the dashed green curves for $\Delta = 250$ meV. Here we set $v_F = 2.55$ eV\AA and $\lambda = 150$ eV\AA$^3$.}
\label{fig:contour}
\end{figure}

The constant energy contour for band dispersion is shown in Fig.~(\ref{fig:contour}). The rashba interaction term referring the first term of Eq.~\ref{eqn:bandham} shows the regular circular shape Fermi surface. This converts into the hexagonal distorted or snowflake shape with the inclusion of warping effect or the term proportional to the third power of the wave-vector that reduces the infinite mirror planes to three. This scenario further show significant effect with time-reversal symmetry breaking finite surface gap. 
Corresponding to the warped and gapped Hamiltonian, the energy eigenstates are
\begin{equation}
\ba
|u^\pm_{\bm k}\rangle &\dps  =\frac{1}{\sqrt{2}}
\left(
\begin{array}{c}
\pm i e^{-i\theta} \sqrt{1\pm \frac{\lambda k^3 \cos 3\theta + \Delta}{\varepsilon_{{\bm k}}}}   \\
\sqrt{1\mp \frac{\lambda k^3 \cos 3\theta + \Delta}{\varepsilon_{{\bm k}}}}
\end{array}
\right).
\ea
\end{equation}
\subsubsection{Induced electric current by thermal field}
First, we will calculate the electric current generated by thermal field ${\bm \nabla} T$ for first-order correction to density matrices i.e. for $n=1$ in Eq.~(\ref{eqn:density}). We consider the gradient field in x-direction ${\bm \nabla}T = \nabla_x T \hat{x}$ and compute the generated electrical current in $\hat{x}$ amd $\hat{y}$ directions using Eq.~(\ref{eqn:currentdef}). Based on Eq.~(\ref{eqn:density}), two driving terms such as field-dependent term and scattering term would give rise to the current. Here we mainly focus on the current production due to the field-dependent driving term that do not include interaction effects. Keeping this in mind, we have
\begin{equation}
\ba
J_{x,Q}^{(1),E} &\dps =-\frac{e}{\hbar} \frac{\nabla_x T}{T} \sum_{\bm k} \bigg\{ \frac{\tau}{\hbar} (\varepsilon_{\bm k}-\mu)\bigg(\pd{\varepsilon_{\bm k}^{+}}{k_x}\bigg)^2 \pd{f_{\bm k}^{0}}{\varepsilon_{\bm k}}\\[3ex]
&\dps + Re\bigg(i (\mathcal{R}_{k_x}^{+-})^2(\varepsilon_{\bm k}^{+}f_{\bm k}^{0,+} - \varepsilon_{\bm k}^{-}f_{\bm k}^{(0),-})\bigg) \bigg\},
\ea
\end{equation}
\begin{equation}
\ba
J_{y,Q}^{(1),E} &\dps =-\frac{e}{\hbar}  \frac{\nabla_x T}{T}  \sum_{\bm k} \bigg\{ \frac{\tau}{\hbar}  (\varepsilon_{\bm k}-\mu)\pd{\varepsilon_{\bm k}^{+}}{k_x}\pd{\varepsilon_{\bm k}^{+}}{k_y} \pd{f_{\bm k}^{0}}{\varepsilon_{\bm k}}\\[2ex]
&\dps + Re\bigg(i \mathcal{R}_{k_y}^{+-}\mathcal{R}_{k_x}^{+-}(\varepsilon_{\bm k}^{+}f_{\bm k}^{0,+} - \varepsilon_{\bm k}^{-}f_{\bm k}^{(0),-})\bigg) \bigg\}.
\ea
\end{equation}
The first part proportional to the first power of transport time $\tau$ generates the intraband contribution and the other part gives the interband contribution due to the involvement of the Berry connection which relates two neighboring bands. 
Considering the modeled Hamiltonian discussed earlier and assuming the linear wave-vector dependent term dominant over other terms in the Hamiltonian, we find that for the longitudinal direction the interband contribution vanishes and only intraband contribution produces finite value to the induced electrical current. 
\begin{equation}
\ba
J_{x,Q}^{(1),E} 
&\dps = \frac{e\tau}{4\pi\hbar^2} \bigg\{ \mu \log 2 + T\frac{\pi^2}{6} \bigg\} \nabla_x T.
\ea
\end{equation} 
In Hall direction, the interband part yields finite $J_{y,Q}^{(1),E}$ while intraband gives zero result.
\begin{equation}
\ba
J_{y,Q}^{(1),E} 
&\dps = \frac{e}{8\pi \hbar} \frac{\Delta \log 2}{\mu}\nabla_x T.
\ea
\end{equation}
This depicts the dependence of the Hall component of induced electric current linearly on the surface gap and inversely to the chemical potential while the longitudinal component comes directly proportional to $\mu$. However, the hexagonal warping does not generate any effect on both the components.\\
For second-order $n=2$ case, the longitudinal component is
\begin{equation}
\ba
J_{x,Q}^{(2),E} &\dps = \frac{e}{32 \pi \hbar} \frac{\Delta \lambda}{v_F^2 T}(\nabla_x T)^2   \bigg(2\log 2 + \frac{\pi^2 T}{12\mu}\bigg).  
\ea
\end{equation}
Similarly the Hall component is
\begin{equation}
\ba
J_{y,Q}^{(2),E} &\dps = \frac{e}{32 \pi \hbar} \frac{\tau \lambda}{\hbar v_F^2 T}(\nabla_x T)^2   \bigg( \mu^2 \log 2 + \frac{\pi^2}{3} T \mu + \frac{9}{2} T^2 \zeta (3) \bigg).  
\ea
\end{equation}
Here, interband yields to the induced electrical longitudinal and intraband to the induced electrical hall current. Also, both contributions gives directly proportional behavior with the first power of the warping coefficient.
 Thus, the warping term show significant contribution in the higher-order case. 
\subsubsection{Thermal current by temperature gradient}
In the same spirit of an electrical case, we have computed the thermal currents in response to the temperature gradient applied in the x-direction. We find that the linear case yields
\begin{equation}
\ba
J_{x,Q}^{(1),Q} &\dps = \frac{\tau}{8\hbar^2} T \mu \frac{\pi}{6} \nabla_x T.
\ea
\end{equation} 
\begin{equation}
\ba
J_{y,Q}^{(1),Q} &\dps = \frac{T}{8 \hbar} \frac{\Delta \pi}{12 \mu}\nabla_x T.
\ea
\end{equation}
Both components show same dependency to the surface gap and chemical potential as the electrical current. In addition to this, these are directly proportional to the first power of the temperature. This dependency stems from the definition of the thermal current, contains $(\varepsilon_{\bm k} - \mu)$ factor. Similarly for non-linear $n=2$ case, the expressions for currents read 
\begin{equation}
\ba
J_{x,Q}^{(2),Q} &\dps = \frac{1}{32 \pi \hbar} \frac{\Delta \lambda}{v_F^2}(\nabla_x T)^2   \bigg(\frac{\pi^2}{6} + \frac{3\zeta (3) T}{2}\bigg).  
\ea
\end{equation}
\begin{equation}
\ba
J_{y,Q}^{(2),Q} &\dps = \frac{1}{32 \pi \hbar} \frac{\tau \lambda}{\hbar v_F^2 }(\nabla_x T)^2   \bigg( \frac{\pi^2 \mu^2}{6} + 9\zeta (3) T \mu + \frac{7\pi^4 T^2}{30} \bigg).  
\ea
\end{equation}

\subsubsection{Electrical current by electric field}
In the application of an external electric field ${\bm E} = E \hat{x}$, the electrical current in $x$ and $y-$-directions corresponding to $n=1$ case will be
\begin{equation}
\ba
J_{x,E}^{(1),E} &\dps =  \frac{e^2\tau}{4\pi\hbar^2} \bigg\{ \frac{\mu}{2} +  T \log 2\bigg\} E_x . \\ [2ex]
J_{y,E}^{(1),E} &\dps =  \frac{e^2}{8\pi \hbar} \frac{\Delta}{2\mu} E_x.
\ea
\end{equation}
Similarly for $n=2$ case,
\begin{equation}
\ba
J_{x,E}^{(2),E} &\dps =  \frac{e^3}{32 \pi \hbar} \frac{\Delta \lambda}{v_F^2 T}E_x^2   \bigg(1 + \frac{T \log 2 }{2\mu}\bigg) \\[2ex]
J_{y,E}^{(2),E} &\dps =  \frac{e^3}{32 \pi \hbar} \frac{\tau \lambda}{\hbar v_F^2 T}E_x^2   \bigg( \frac{\mu^2}{2} + 2 T \mu \log 2 + \frac{\pi^2 T^2}{6} \bigg).
\ea
\end{equation}
Using these expressions, we analyze the non-linearity in the Seebeck effect and the Pelteir effect in terms of the ratio of non-linear to the linear conductivities in the next section. 
\section{Numerical Analysis}
\begin{figure}[htbp]
\centering
\subfigure[noonleline][]
{\label{fig:ratioAxW}\includegraphics[width=7cm, height= 5cm]{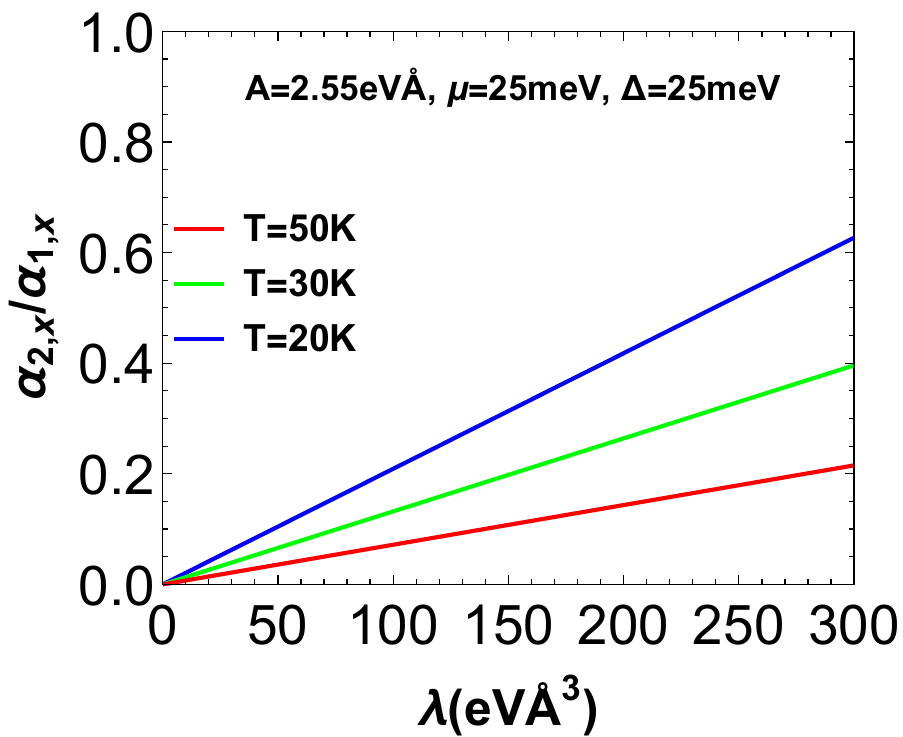}}
\subfigure[noonleline][]
{\label{fig:ratioAyW}\includegraphics[width=7cm, height= 5cm]{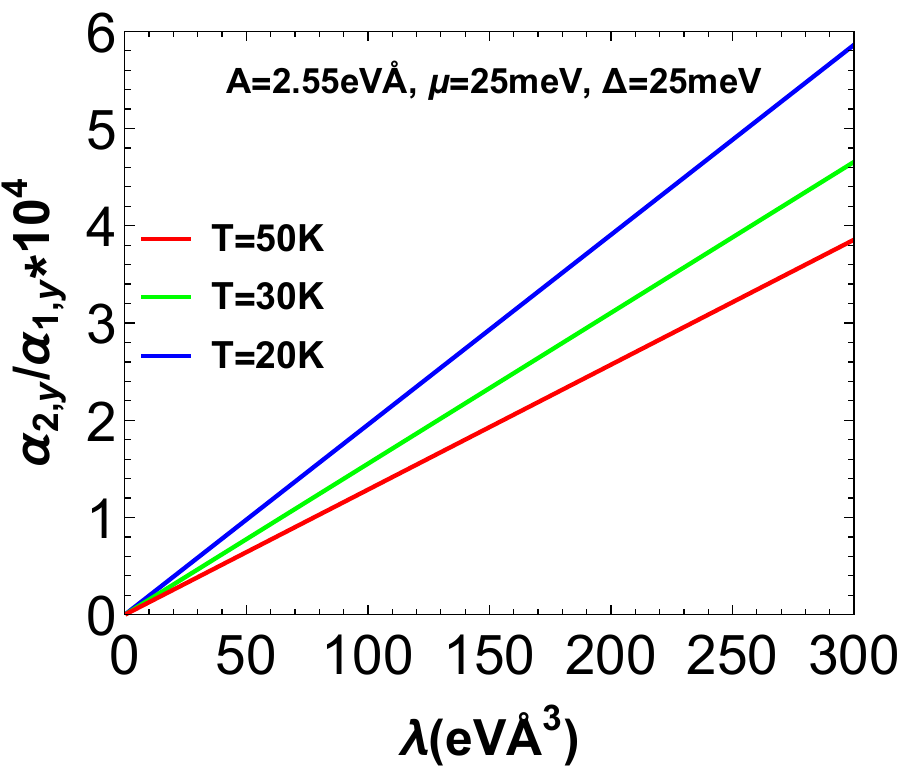}}
\caption{Variation of the ratio of non-linear to linear thermoelectric conductivities with the warping coefficient at different temperature values. (a). Longitudinal case; (b). Hall case.}
\label{fig:RatioAW}
\end{figure}
In Fig.~\ref{fig:RatioAW}, we show the results for the ratio of non-linear conductivity to the linear conductivity $\alpha_{2}/\alpha_{1}$, as a function of warping coefficient $\lambda$ (eV\AA$^3$) for different low temperature values. Here we keep the surface gap $\Delta = 25$ meV, chemical potential $\mu = 25$ meV and $v_F = 2.55$ eV\AA. We find that the ratio increases linearly with the rise of warping coefficient. However, with the rise of temperature, the magnitude of the conductivities ratio decreases due to the consideration of only intrinsic contribution or the non-disorder case. On comparing the longitudinal and Hall cases, we find that the Hall case contributes more to the variation of the non-linear Seebeck and Peltier coefficient. We see that both the ratios gives first-order dependence with the strength of hexagonal degree of distortion. 
\begin{figure}[htbp]
\centering
\subfigure[noonleline][]
{\label{fig:ratioAxG}\includegraphics[width=7cm, height= 5cm]{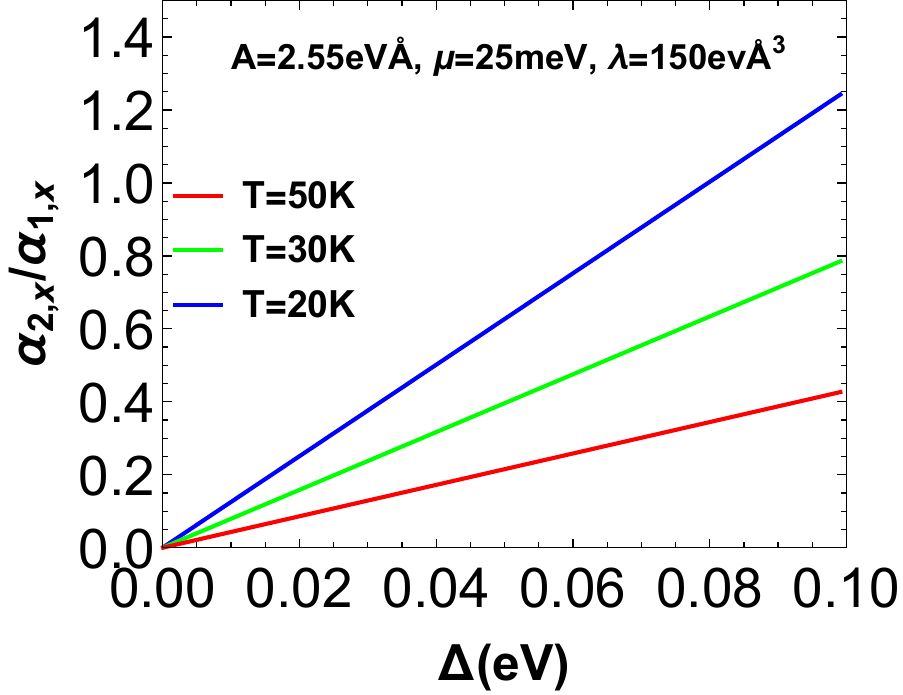}}
\subfigure[noonleline][]
{\label{fig:ratioAyG}\includegraphics[width=7cm, height= 5cm]{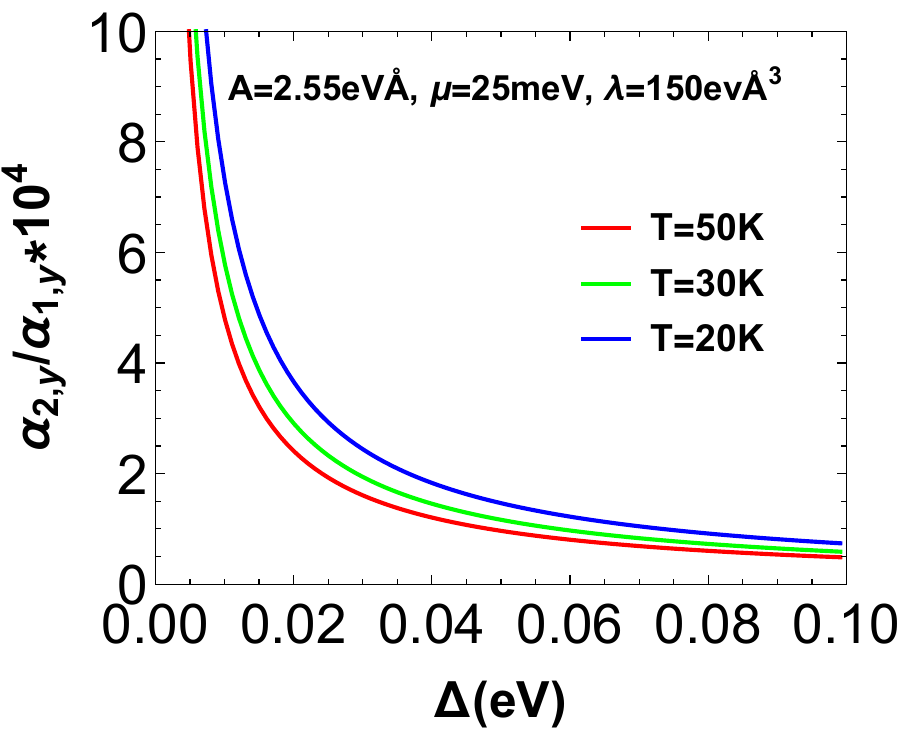}}
\caption{Plot for the ratio of second-order thermoelectric conductivity to the linear one with the gap at different temperature values. (a). Longitudinal case; (b). Hall case.}
\label{fig:RatioAG}
\end{figure}

In Fig.~\ref{fig:RatioAG}, the influence of the surface gap $\Delta$ on $\alpha_{1}/\alpha_{2}$ keeping $\lambda = 150$ eV\AA$^3$, $\mu = 25$meV and $v_F = 2.55$ eV\AA is shown. Results indicate that the larger the surface gap is, stronger the longitudinal ratio is at all temperature values. This variation is also consistent with the analytical results. Contrary to it, the Hall ratio decreases with the rise of gap at all temperatures. Further with the variation of chemical potential, we find the exponential decay in the ratio of conductivities with the temperature. However there is an increase in the Hall ratio as chemical potential $\mu$ increases as shown in Fig.~\ref{fig:RatioAT}. Similar behavior can be seen in the ratio of second-order to the first-order thermal conductivities for different parameters. Here for demonstration we have only shown the variations of thermal conductivities ratio numerically. \\

\begin{figure}[htbp]
\centering
\subfigure[noonleline][]
{\label{fig:ratioAxT}\includegraphics[width=7cm, height= 5cm]{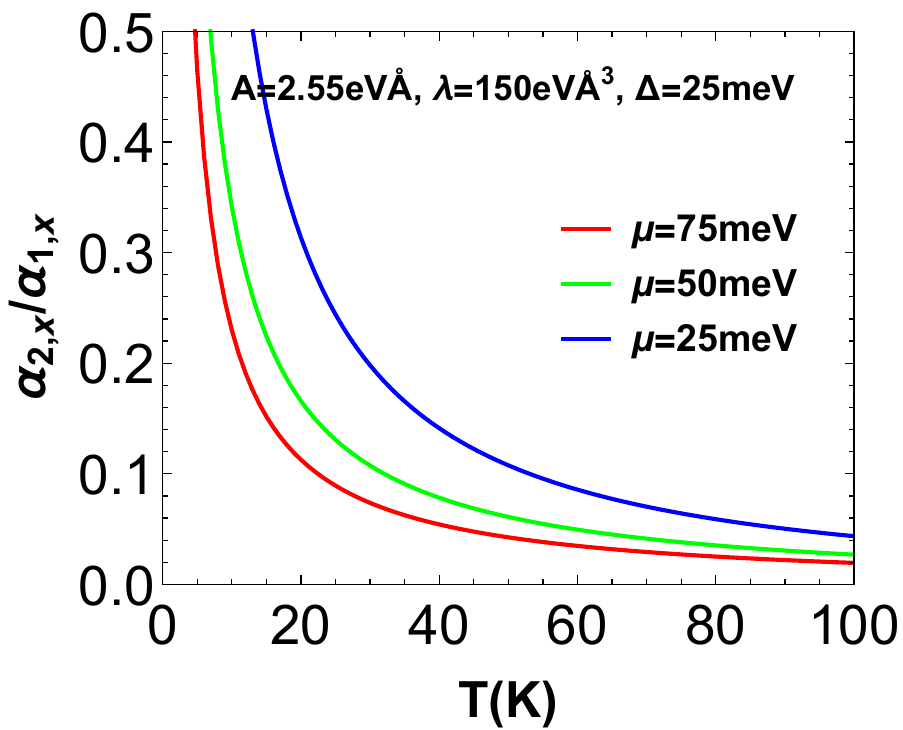}}
\subfigure[noonleline][]
{\label{fig:ratioAyT}\includegraphics[width=7cm, height= 5cm]{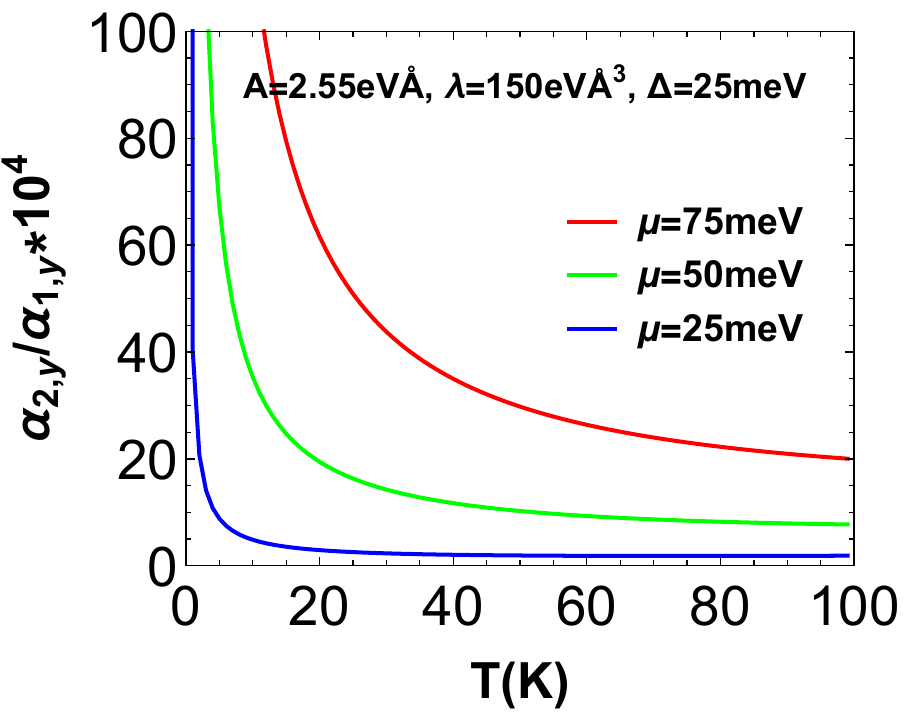}}
\caption{The ratio of second-order response to the linear one with the temperature at different chemical potential values. (a). Longitudinal case; (b). Hall case.}
\label{fig:RatioAT}
\end{figure}
The above analysis indicate that the non-linear response or the non-linear thermoelectric effects arise due to the distorted Fermi surface by the hexagonal warping effect. If $\lambda =0$ eV\AA$^3$, interband and intraband second-order response vanishes, thus the conductivities ratio approaches to zero. In addition to this, the surface gap shifts the Fermi surface from its original position which influences the conductivity. This has been observed in Fig.~\ref{fig:RatioAG}. In the linear regime, the Hall conductivity varies linearly with the gap and the longitudinal shows independent behavior with the surface gap. However, the non-linear conductivity shows opposite behavior, means $\alpha_{2,x} \propto \Delta$ and $\alpha_{2,y}$ shows no variation with $\Delta$. Further, the ratio shows turnover when $\Delta$ becomes larger than the chemical potential $\mu$ as shown in Fg.~\ref{fig:RatioAG}.\\
The non-linear Seebeck and Peltier coefficients, proportional to the ratio of the non-linear to the linear responses (Eq.~\ref{eqn:SE}and \ref{eqn:PE}) increases with gap when the field is applied in the x-direction and the current is also measured in the same direction. However, it reduces in the perpendicular direction (y-direction). But warping enhances the non-linearity contribution in both directions. \\
The total conducitivity, a sum of the interband and intraband contributions is presented here in detail. Intraband processes lead to finite longitudinal response to the first-order in an external field whether electric field or temperature gradient is applied. Howver, interband gives Hall contribution. To the next-order in field, the intraband yields Hall conductivity and interband produces longitudinal one. Thus, to analyze the non-linear effects one requires both inter and intaband participation. One important message from this analysis is the finite contribution to the non-linear effect only if we consider a non-zero warping and gap effects in picture. To the zero case, longitudinal case vanishes and Hall one diverges. Hence, the both surface and warping effects are pivotal to understand the non-linear effects. However, with the longitudinal geometry one can get more deviation from the linear regime with these terms and enhance the overall magnitude of the thermoelectric effects.\\
We have only considered the intrinsic contribution means the contribution stems from Berry connection and field driving terms. Disorder contribution will come from the scattering term $J(f_{\bm k})$ as mentioned in the theoretical description. Here one requires the different interactions effects such as electron-impurity, electron-phonon, etc. and these will play a significant role at temperature behavior of the responses. These considerations are beyond the scope of the present study.







\section{Conclusion}
Beyond the linear response regime, we have studied the thermoelectric properties of a topological insulator. Our formalism is based on the quantum kinetic approach and includes both intrinsic and extrinsic responses to the electric and thermal fields. However, we mainly focus on the intrinsic part of the nonlinear effects. We evaluate the thermal and electric currents expressions to the second power of the external field. We have found that the nonlinearity arises due to the ratio of the responses from interband and intraband processes. Thus, both the contributions play significant role to understand the nonlinear Seebeck and Peltier effects. Furthermore, we analyzed the warping and the surface gap effect on these properties using this formalism and provides fundamental importance to the nonlinear effects in topological insulators which will be beneficial to improve the technological advances.  

\acknowledgments

PB acknowledges the National Key Research and Development Program of China grant No. 2017YFA0303400 and NSAF China grant No. U1930402 for financial support. 

%

\appendix
\section{Derivation of Eq.~\ref{eqn:density}}
For the diagonal case, the kinetic equation reads
\begin{equation}
\ba
 &\dps \frac{d f_{d,{\bm k}}^{(n)} }{dt} + \frac{f_{d,{\bm k}}}{\tau} = \mathcal{D}_T (f_{d,{\bm k}}^{(n-1)}) - J_d[f_{\text{od},\textbf{k}}^{(n)}].
\ea
\end{equation}
Solving this equation using the integrating factor, we have
\begin{equation}
\ba
f_{d,{\bm k}}^{(n)}  &\dps = \int_{0}^{\infty} dt' e^{-t'/\tau} \bigg[ \mathcal{D}_T (f_{d,{\bm k}}^{(n-1)}) - J_d[f_{\text{od},\textbf{k}}^{(n)}] \bigg].
\ea
\end{equation}
\begin{equation}
\ba
f_{d,{\bm k}}^{(n)}&\dps = \tau  \bigg[ \mathcal{D}_T (f_{d,{\bm k}}^{(n-1)}) - J_d[f_{\text{od},\textbf{k}}^{(n)}] \bigg].
\ea
\end{equation}
\begin{equation}
\ba
f_{d,{\bm k}}^{(n)} &\dps = \tau  \bigg[\frac{1}{\hbar} \frac{{\bm \nabla} T }{T}\cdot \pd{f_{d,{\bm k}}^{(n-1)}}{{\bm k}} - J_d[f_{\text{od},\textbf{k}}^{(n)}] \bigg].
\ea
\end{equation}
Similarly, for the off-diagonal case
\begin{equation}
\ba
&\dps \frac{d f_{\text{od},{\bm k}}^{(n)}}{dt} + \frac{i}{\hbar} [H_0, f_{\text{od},{\bm k}}^{(n)}] +  \frac{f_{\text{od},{\bm k}}}{\tau} = \mathcal{D}_{T}(f_{\text{od},{\bm k}}^{(n-1)}) - J_{\text{od}}[f_{\text{d},{\bm k}}^{(n)}].
\ea
\end{equation}
\begin{equation}
\ba
f_{\text{od},{\bm k}}^{(n)} &\dps = \int_{0}^{\infty} dt' e^{- t'/\tau} e^{-i\varepsilon_{m{\bm k}}t'/\hbar} \bigg[ \frac{1}{\hbar}\frac{{\bm \nabla}T}{T} \cdot \bigg( \pd{f_{\text{od},{\bm k}}^{(n-1)}}{{\bm k}}  \\[2ex]
&\dps \quad\quad - i [\mathcal{R}_{\bm k}, f_{{\bm k}}^{(n-1)}]^{mm'}\bigg) - J_{\text{od}}[f_{\text{d},{\bm k}}^{(n)}]  \bigg]  e^{i\varepsilon_{m'{\bm k}}t'/\hbar}. 
\ea
\end{equation}
\begin{equation}
\ba
f_{\text{od},{\bm k}}^{(n)} &\dps = \frac{\hbar}{i(\varepsilon_{m{\bm k}} - \varepsilon_{m'{\bm k}})}\bigg[ \frac{1}{\hbar}\frac{{\bm \nabla}T}{T} \cdot \bigg( \pd{f_{\text{od},{\bm k}}^{(n-1)}}{{\bm k}}  \\[2ex]
&\dps \quad\quad - i [\mathcal{R}_{\bm k}, f_{{\bm k}}^{(n-1)}]^{mm'}\bigg) - J_{\text{od}}[f_{\text{d},{\bm k}}^{(n)}]  \bigg] .
\ea
\end{equation}
Here the commutation relation between $\mathcal{R}_{\bm k}$ and $f_{\bm k}$ is
\begin{equation}
\ba
[\mathcal{R}_{\bm k}, f_{\bm k}^{(n)}]^{mm'} &\dps= \sum_{m''} \langle m \vert \mathcal{R}_{\bm k} \vert m'' \rangle \langle m'' \vert f_{{\bm k}} \vert m' \rangle \\[2ex]
&\dps \quad\quad - \langle m \vert f_{{\bm k}} \vert m'' \rangle \langle m'' \vert \mathcal{R}_{\bm k} \vert m' \rangle.
\ea
\end{equation}

\end{document}